\documentclass[12pt]{article}
\usepackage{amssymb,epsf}
\def\lsim{\mathrel{\rlap {\raise.5ex\hbox{$ < $}}
{\lower.5ex\hbox{$\sim$}}}}
\def\gsim{\mathrel{\rlap {\raise.5ex\hbox{$ > $}}
{\lower.5ex\hbox{$\sim$}}}}
\def\np#1#2#3{Nucl. Phys. {\bf{B#1}} (#2) #3}
\def\pl#1#2#3{Phys. Lett. {\bf{B#1}} (#2) #3}

\begin{document}
\renewcommand{\theequation}{\arabic{section}.\arabic{equation}}
\begin{titlepage}
\begin{flushright}
CERN-TH/98-352 \\
NEIP-98-015\\
hep-th/9811096 \\
\end{flushright}
\begin{centering}
\vspace{.3in}
{\Large \bf Non-perturbative corrections
in $N=2$ strings $^\dagger$}\\
\vspace{.15in}
\vspace{1.2cm}
{Andrea GREGORI}\\
\vspace{1cm}
{\it Theory Division, CERN $^1$}\\
{\it 1211 Geneva 23, Switzerland}\\
{\it and}\\
{\it Institut de Physique Th\'{e}orique,
Universit\'{e} de Neuch\^{a}tel}, \\
{\it 2000 Neuch\^{a}tel, Switzerland}\\
\vspace{1.8cm}
{\bf Abstract}\\
\end{centering}
\vspace{.5cm}
We investigate the non-perturbative equivalence of some heterotic/type IIA
dual pairs with $N=2$ supersymmetry.
We compute $R^2$-like corrections, both on the type IIA and on the
heterotic side. The coincidence of
their perturbative part provides a test of duality.
The type IIA result is then used to predict the full, non-perturbative
correction to the heterotic effective action.
We determine the instanton numbers and the Olive--Montonen duality groups.
\vspace{2.7cm}
\begin{flushleft}
CERN-TH/98-352 \\
NEIP-98-015 \\
November 1998 \\
\end{flushleft}
\hrule width 6.7cm
\vskip.1mm{\small \small \small
$^\dagger$\ Talk given at the {\sl 6th Hellenic School and Workshops
on Elementary Particle Physics}, Corfu,6--26 September 1998. To appear in the
proceedings. Research supported in part by the EEC under the
TMR contract ERBFMRX-CT96-0045 and by th Swiss Office
for Education and Science (ofes 95.0856).\\
$^1$ E-mail address: agregori@mail.cern.ch.}
\end{titlepage}
\newpage

All known different string constructions are conjectured to be part of 
a unique theory. This means in particular that compactifications
with different field-content and supersymmetry correspond to
different regions of a unique moduli space.
The duality conjecture \cite{ht}--\cite{vw2}
has received much attention in the past years,
and has passed many tests, performed 
on different compactifications, for various dimensions and number of 
supersymmetries. 
However, in spite of this, very few tests have been performed
by using string theory \cite{kv,re}: 
most of them rely on classical arguments and 
on the properties of the supergravity effective theories, 
or on the geometry of the classical moduli space.
The aim of our work is to provide a test of the duality
for some specific string dual pairs, constructed at a point in the moduli 
space in which it is possible to explicitly solve the two-dimensional
conformal theory and make one-loop string computations.
The models we consider are type IIA/heterotic string pairs with
$N=2$ supersymmetry in four space-time dimensions.
The starting point is the equivalence of all the $N=4$ supergravity 
theories, and the (well-supported but still only conjectured)
equivalence of the overlying string theories \cite{ht}.   
The Hull--Townsend duality states the equivalence of the
type IIA string compactified on $T^2 \times K3$ and the heterotic string
compactified on the torus $T^6$.
In both theories, the space of the moduli-field vacuum expectation values
is spanned by 134 physical scalars, which are coordinates of the coset space
\cite{ht,fk}:
\begin{equation}
\left( {SL(2,R) \over U(1)} \right)_S \times 
\left( {SO(6,6+r) \over SO(6) \times SO(6+r) }\right)_T~,~~~~r=16~.
\end{equation}
On the heterotic side the dilaton $S_{\rm Het}=S$ is in the gravitational
multiplet, while on the type IIA side it is one of the moduli of the vector 
multiplet: $S_{\rm II}=T^1$, $T^1$ being the volume form of the two-torus.
The duality therefore is non-perturbative, involving an interchange of
fields of the $S$ and $T$ manifolds \cite{ht}.
More $N=4$ dual pairs, which involve the same 
field exchange, can be constructed 
by going to the $T^4 /Z_2$ orbifold limit of K3
and projecting out some states of the $r=16$ theory
with particular ``freely-acting'' projections, which
remove some of the orbifold fixed points, without breaking supersymmetry
further \cite{vw,6auth}. 
In this way, the $r=8$ and $r=4$ dual pairs can be constructed.
Dual pairs with $N=2$ supersymmetry are then obtained by breaking the $N=4$
supersymmetry with a further $Z_2^{\rm f}$ projection, which still has
the property of acting freely \cite{gkp}. 
The action is then a rotation on some 
coordinates accompanied by a translation in others.
This implies that the supersymmetries are broken ``spontaneously'',
and it is possible to restore them by going to a specific corner in the 
moduli space of these orbifolds \cite{kk}--\cite{kkprn}.
The theories with $N=4$ and $N=2$ are then continuously related, and the 
duality of the $N=4$ phase is naturally inherited by the $N=2$ phase.
The perturbative heterotic scalar manifolds of the models we consider
are
\begin{equation}
{SU(1,1) \over U(1)} \times {SO(2,2+N_V) \over SO(2) \times SO(2+N_V)}
~~~{\rm and}~~~{SO(4,4+N_H) \over SO(4) \times SO(4+N_H)}~,
\end{equation}
with $N_V=N_H=r=8,4,2$,
for moduli respectively in the vector multiplets and hypermultiplets.

\vspace{.4cm}
\noindent
{\sl Type IIA}
\vspace{.4cm}

On the type IIA side, these models correspond to self-mirror Calabi--Yau
threefolds with Hodge numbers $h^{1,1}=N_V+3=h^{2,1}=N_H+3$, which are
K3 fibrations. This last property is a necessary condition for
the existence of the heterotic duals \cite{klm,al}, and is directly
related to the spontaneous breaking of the $N=4$ supersymmetry
\cite{gkp,gkpr}. 
The gauge group is Abelian:
\begin{equation}
G=\left[U(1)^2 \right] \times U(1)^r
\end{equation}
(within square brackets, we indicate the gauge symmetry
corresponding to the bosons of the untwisted sector of the orbifold).
The $r=8$ model is obtained from the $N=4$, $r=16$,
with the only SUSY-breaking projection $Z_2^{\rm f}$.
The $r=4$ and $r=2$ orbifolds are then obtained by applying
respectively one and two ``$Z_2^D$''-projections; these latter
act on the $Z_2$ orbifolds by 
modding out the states that are odd under the $D$-symmetries,
which exchange twist fields $\sigma_{\pm}$ and project
the untwisted vacua $V_{nm}$ \cite{dvv}: 
\begin{equation}  
D:~~\left(\sigma_+,\sigma_-,V_{nm} \right) \rightarrow
\left(\sigma_-,\sigma_+,(-)^{nm} V_{nm} \right)~.
\end{equation}

\vspace{.4cm}
\noindent
{\sl Heterotic}
\vspace{.4cm}

The heterotic duals are constructed by following analogous steps,
starting from the $N=4$, ``$E_8 \times E_8$''
string compactified on $T^6$.
However, on the heterotic side, in order to obtain the same
spectrum as on the type IIA side, we need to introduce also
$Z_2$-projections (discrete Wilson lines), which break
the initial gauge group to
$\left[ U(1)^2 \right] \times U(1)^{16}$.
The heterotic analogue of the $Z_2^{\rm f}$-projection
acts on the coordinates of the compact space $T^6=T^2 \times T^4$
as a rotation on $T^4$ accompanied by a translation in $T^2$.
On the $c=(0,16)$ part of the currents, it acts instead
as a lattice exchange \footnote{This construction is similar
to that of \cite{fhsv}. For the details, see \cite{gkp}.}.
The result of this operation is the 
spontaneous breaking of the $N=4$ supersymmetry to $N=2$
and a reduction of the rank of the gauge group from $r=16$
(we omit for simplicity the gauge group of the torus)
to $r=8$. Out of the Abelian point, the rank $r$ part
of the gauge group is realized at the level 2.
The models with $r=4$ and $r=2$ are then constructed by applying
further projections, the heterotic analogue of the 
$Z_2^{D}$-projections, which act on the
currents as a further lattice exchange, and as a
translation on some specific directions of $T^4$.
Out of the $U(1)^r$ point, the gauge group is realized
at the level $16/r$.
The action of the translations associated to the above projections
is not completely arbitrary, being constrained by modular invariance.
There is, however, some freedom in choosing the specific embedding
of those, and it turns out that there exists a particular choice
that corresponds to the same region of the perturbative moduli space
as that selected in the type IIA duals. This renders possible
the identification of the map between the moduli of the 
heterotic and type IIA dual constructions (for details, see \cite{gkp}).

\vspace{.4cm}
\noindent
{\sl The gravitational corrections}
\vspace{.4cm}

In the particular region of moduli space we are considering, 
it is possible to construct the partition function of these orbifolds
and to make a direct computation of the
string corrections to any term of the effective action.
The particular term we consider here is a well-defined combination of
the $R^2 = \langle R_{\alpha \beta \gamma \delta}
R^{\alpha \beta \gamma \delta} \rangle$ gravitational 
amplitude \cite{fg}--\cite{hmn=2} and of gauge amplitudes,
namely the $F_{\mu \nu}F^{\mu \nu}$ amplitude corresponding to the gauge 
bosons of the $U(1) \times U(1)$ symmetry originating from the heterotic 
untwisted torus $T^2$ (or, equivalently, on the 
type IIA side, from the untwisted 
sector) and the amplitude corresponding to the gauge bosons of the rank $r$
(level $16/r$) part. The $F_{\mu \nu}F^{\mu \nu}$ terms, although vanishing
at a generic point in the moduli space, turn out to be
necessary for a comparison of the type IIA and heterotic theories.
These amplitudes are computed by inserting specific operators in the 
partition function.

On the type IIA side, the correction to the $R^2$ term is
computed at one loop and is
obtained by inserting the operator $2 Q^2  \overline{Q}^2$,
$Q$ and $\overline{Q}$ being respectively the left-handed
and right-handed helicity operators. Owing to the
world-sheet symmetry between left and right movers, the insertion of the
helicity operators amounts eventually to a scalar derivative
on modular forms, considered as functions of
the modular parameter $\tau$ of the world-sheet torus.  
This amplitude depends on the moduli $T^1$, $T^2$, $T^3$
(the K\"{a}hler classes of the three tori of the compact space) and,
owing to the absence of $\Delta N_V \neq 0$ and/or $\Delta N_H \neq 0$
singularities, it is regular. 
The gauge amplitudes, on the other hand,
vanish trivially, being the gauge-bosons Ramond--Ramond states.
After specifying the direction of the $Z_2^{\rm f}$ and
$Z_2^D$ translations, which we choose to act on the momenta,
we obtain \footnote{For details, see 
\cite{6auth,solving,kkprn,dkl,mast,lest,hmbps}.}:
\begin{eqnarray}
{16 \, \pi^2 \over g^2_{\rm grav} \left( \mu^{(\rm II)}\right)} & = &
-{3 N_V \over 4} \log {\mu^{(\rm II)} {\rm Im} T^1 \over M^{(\rm II)}}
\left\vert \eta \left( T^1 \right)  \right\vert^4
- \left( 2-{N_V \over 4}  \right) 
\log{\mu^{(\rm II)} {\rm Im} T^1 \over M^{(\rm II)}}
\left\vert \vartheta_4 \left( T^1 \right)  \right\vert^4 \nonumber \\
&& -2 \log{\mu^{(\rm II)} {\rm Im} T^2 \over M^{(\rm II)}}
\left\vert \vartheta_4 \left( T^2 \right)  \right\vert^4 
-2 \log{\mu^{(\rm II)} {\rm Im} T^3 \over M^{(\rm II)}}
\left\vert \vartheta_4 \left( T^3 \right)  \right\vert^4~
+{\rm const.}~, 
\label{thr}
\end{eqnarray}
with $\mu^{(\rm II)}$ and $M^{(\rm II)}$ the
infrared cut-off and the string scale of type IIA, respectively.

On the heterotic side, both genus-0 and genus-1 contribute
to the $R^2$ term. However, the genus-1
contribution, which depends on the moduli $T$ and $U$, respectively
the K\"{a}hler class and the complex structure modulus of the untwisted 
torus $T^2$, is not regular as a function of these fields.
In fact, its beta-function jumps at the special values of these moduli
for which new vector- or new hyper-multiplets appear in the spectrum
(in particular, this happens when the $U(1)$ factors from the torus
are enhanced to $SU(2)$).
Moreover, on the heterotic side there is no world-sheet symmetry
between left and right movers, and the insertion of the ``gravitational''
operator in the partition function does not amount, after saturation of
the fermion zero-modes, to a scalar but to a covariant
derivative on modular forms \cite{hmbps}--\cite{gauge}:
\begin{equation}
{\cal D}_{\tau}=\partial_{\tau}-{id \over 2 {\rm Im} \tau}~,
\end{equation}
where $d$ is the weight of the form.
In order to obtain a regular amplitude, we correct the $R^2$ 
term by adding to it a term proportional to
the $F^2$ amplitude of the torus, whose beta-function
vanishes for generic values of the moduli $T$ and $U$, but
jumps at the singular points, in a way opposite to the jumping
of the $R^2$ beta-function. This amplitude is now regular, but it still 
corresponds to an operator that acts as a covariant derivative.
In order to obtain a scalar operator, we add also a term proportional to the
$F^2$ amplitude of the level $16/r$ part of the gauge group.
The latter has vanishing beta-function and is regular, for any value of the 
moduli $T$ and $U$. The proper combination, which has the required
properties of regularity and ``holomorphicity'' (i.e. no 
${\rm Im} \tau$-covariantization), is therefore:
\begin{equation}
Q^2_{\rm grav}= R^2 + 
{1 \over 12} \left( F_{\mu \nu} F^{\mu \nu}  \right)_{T^2}
+{5 \over 36 N_V} \left( F_{\mu \nu} F^{\mu \nu}  \right)_{G_{(N_V)}}~.
\label{q2}
\end{equation}
Saturation of the fermion zero-modes implies that only the
so-called $N=2$ sector of the heterotic orbifolds contributes to
the one-loop amplitude. This sector corresponds to the terms of the
partition function that are projected and twisted by $Z_2^{\rm f}$.
A remarkable fact is that this sector is universal: namely, it is
the same for all three models, and it does not depend on the
extension of the rank $r$ part of the gauge group, $G_{(N_V)}$: 
$U(1)^r$, $\left( SU(2)\vert_{16/r} \right)^r$, etc.
As a consequence, the one-loop amplitude of (\ref{q2}) is the same
for all models \cite{gkp}. 
Putting together the tree-level and the one-loop contributions,
we get, for all the models:
\begin{eqnarray}
{16 \, \pi^2 \over g^2_{\rm grav} \left(  \mu^{(\rm Het)} \right)} & = &
16 \, \pi^2 {\rm Im}S -2 \log {\rm Im} T 
\left\vert \vartheta_4 \left( T \right)  \right\vert^4
-2 \log {\rm Im} U 
\left\vert \vartheta_4 \left( U \right)  \right\vert^4 \nonumber \\
&& +4 \log {M^{(\rm Het)} \over \mu^{(\rm Het)}} + {\rm const.}~,
\label{hetthr}
\end{eqnarray}
where $S$ is the dilaton--axion field 
\begin{equation}
{\rm Im}S={1 \over g^2_{\rm Het}}~,
\end{equation} 
and we expressed the infrared running in terms of the heterotic
string scale $M^{(\rm Het)} \equiv 1/\sqrt{\alpha'_{\rm Het}}$ 
and cut-off $\mu^{(\rm Het)}$. 
The coefficient $2$ of the terms containing the contribution of the moduli
$T$, $U$ is actually the ``gravitational beta-function'', 
$b_{\rm grav}= (24-N_V+N_H) \big/ 12$.
A comparison of (\ref{thr}) and (\ref{hetthr}) in the heterotic perturbative 
limit, ${\rm Im}S \to \infty$, leads to the identification of 
$T^1$ with $16 \tau_S \big/ N_V$ ($\tau_S=4 \pi S$). This is consistent with 
the interpretation of the compact spaces in the type IIA models
as orbifold limits of K3 fibrations, with base respectively
$T^2 \big/ Z^{\rm f}_2={\bf P}^1$, ${\bf P}^1 \big/ Z_2^D$
and ${\bf P}^1 \big/ (Z_2^D \times Z_2^{D'})$ for $N_V=8,4,2$.
The heterotic ``dilaton'' $\tau_S$ is then the volume form of the base
of these fibrations. 
Moreover, the perturbative corrections, as a function of moduli $T^2$
and $T^3$ on the type IIA side and as a function of $T$ and $U$ on the
heterotic side, 
coincide with the identification of $T^2$, $T^3$ with $T$, $U$. 
These identifications provide a test of the type IIA/heterotic 
duality.

Once established the precise duality map, we use the type IIA result
to predict the full, perturbative and non-perturbative
correction to the heterotic effective action. This is therefore given
by the expression (\ref{thr}), with the above-mentioned substitutions of the
moduli of the type IIA compactification with their heterotic duals.
The contribution of moduli $T$, $U$ to the effective coupling constant,
$16 \, \pi^2 \Big/ g^2_{\rm grav}$, is then universal and coincides
with the one computed at one loop, while
the dilaton dependence, for the various models, is:
\begin{eqnarray}
&& -{3 N_V \over 4} \log \left\vert \eta \left( {16 \tau_S \over N_V}  
\right)  \right\vert^4 -\left( 2-{N_V \over 4} \right) \log 
\left\vert \vartheta_4 \left( {16 \tau_S \over N_V} \right)  \right\vert^4 
\nonumber \\
&& -\left({B_4-B_2 \over 3} -2 b_{\rm grav}  \right)
\log {\rm Im} \tau_S +
{B_4-B_2 \over 3} \log {M_{\rm Planck} \over \mu}~,
\label{dilthr}
\end{eqnarray}
where we have expressed the infrared running in terms of the
Planck mass and the physical cut-off $\mu$, related to the string scales
and cut-offs by:
\begin{equation}
{M^{(\rm II)} \over \mu^{(\rm II)}}=
{M^{(\rm Het)} \over \mu^{(\rm Het)}}=
{M_{\rm Planck} \over \mu}~. 
\end{equation}
The coefficients of the terms in the second line of (\ref{dilthr})
are given in terms of the massless contribution to the
helicity supertraces 
\footnote{For notation and conventions see for instance 
\cite{6auth,kk,bk,n=3}.}
$B_2={\rm Str}(Q+\overline{Q})^2$,
$B_4={\rm Str}(Q+\overline{Q})^4$, and the gravitational
beta-function. Since the amplitude of $Q^2_{\rm grav}$ is regular
in the $(T,U)$-plane, for these quantities
the values have to be taken at a generic point in the moduli space.

The dilaton contribution (\ref{dilthr}) can be interpreted as
a series that contains the perturbative (linear) contribution
and the exponential, instantons contributions. The instantons are due to
Euclidean five-branes also wrapped around the time direction
\cite{hmn=4}. The instanton numbers $k$ are the powers in the expansion
of this expression in the parameter $q_S=\exp 2 \pi i \tau_S$.
We obtain $k=2n$, $n \in N$, for $N_V=8,4$ and $k=4n$ for $N_V=2$.
As a consequence of the non-perturbative 
$Z_2^{\rm f}$ and $Z_2^D$ translations in the plane of 
the dilaton, also the Montonen--Olive $SL(2,Z)_S$-duality group
\cite{s} is broken. The unbroken subgroup depends on $N_V$:
it is a $\Gamma(2)$ subgroup for $N_V=8$,
$\Gamma(8)$ for $N_V=4$, and $\Gamma(16)$ for $N_V=2$.

\vskip 0.3cm
\centerline{\bf Acknowledgements}

\noindent
These notes are based on work done in collaboration with
C. Kounnas and P.M. Petropoulos. 
This research was supported by the Swiss National Science Foundation. 
I would like to thank the organizers
of the ``6th Hellenic School and Workshops on Elementary Particle Physics''.

\end{document}